# Predicting Wrist Osteoporosis from excised human finger bones using spatially offset Raman spectroscopy – A Cadaveric Study

MOHAMMAD HOSSEINI,[1,*] SADIA AFRIN,[2] ANTHONY YOSICK,[2,3] EMMA SCHENKER,[1,4] HANI AWAD,[2,3] AND ANDREW J. BERGER[1,2,*]

[1]*The Institute of Optics, University of Rochester, 275 Hutchison Rd, Rochester, NY 14620, USA*

[2]*Department of Biomedical Engineering, University of Rochester, 207 Robert B. Goergen Hall, Rochester, NY 14620, USA*

[3]*The Center for Musculoskeletal Research, University of Rochester Medical Center, 601 Elmwood Avenue,*
*Box 665, Rochester, NY 14642, USA*

[4]*Department of Physics, University of Florida, Gainesville, FL 32611, USA*

*shossei3@ur.rochester.edu, andrew.berger@rochester.edu

**Abstract:** Osteoporosis and osteopenia remain vastly underdiagnosed. Current clinical screening relies almost exclusively on dual-energy X-ray absorptiometry (DXA), which measures bone mineral density (BMD) but fails to capture the compositional changes that lead to BMD loss. We investigated whether Spatially Offset Raman Spectroscopy (SORS) applied to excised finger bones can assess subsurface biochemical markers capable of diagnosing osteoporosis and osteopenia and predicting wrist DXA T-scores. Raman spectra were acquired *ex vivo* on the mid-shaft of the proximal phalanx of the second digit from 25 female cadavers spanning the three T-score categories (n=8 normal, n=6 osteopenic, and n=11 osteoporotic) at spatial offsets of 0, 3, and 6 mm from a laser irradiation spot. After normalizing spectra to the $PO_4^{3-}$ peak, group-averaged spectra of the three categories, measured at 3-mm offset, showed clear differences in the $CO_3^{2-}$, Amide III, $CH_2$, and Amide I bands. Quantitatively, four out of five mineral-to-matrix ratios differed significantly ($p \leq 0.05$) between normal and osteopenic bone, and between osteopenic and osteoporotic bone, and all five ratios showed significant differences between normal and osteoporotic bone. In contrast, the 0-mm offset suffered diminished contrast, and the 6-mm offset did not enhance discrimination between different groups, compared with the 3-mm offset. A leave-one-out, partial-least-squares regression model built from the 3-mm spectra predicted distal radius DXA T-score with a Pearson correlation of $r = 0.85$ and a root-mean-square error of cross-validation of ≈1 T-score units, correctly classifying 92% of specimens. These results demonstrate that a SORS offset provides sensitivity to biochemical changes and that phalangeal Raman spectra might serve as non-ionizing surrogates for DXA metrics of the wrist. The findings lay the groundwork for translating subsurface Raman spectroscopy into a fast, non-invasive prescreening tool for osteoporosis.

## 1. Introduction

Bone health is a critical public health concern, with osteoporosis and osteopenia affecting over 200 million individuals worldwide, leading to increased fracture risk, significant morbidity, and substantial healthcare costs [1]. Osteoporosis and osteopenia, characterized by reduced bone mass and deteriorated microarchitecture, are traditionally diagnosed using dual-energy X-ray absorptiometry (DXA), which measures bone mineral density (BMD). Bone loss is quantified through a T-score, the number of standard deviations that BMD differs from a normal, healthy individual of the same sex. The World Health Organization (WHO) has established classifications of bone loss severity as Normal (N, T-score > -1), Osteopenia (OPE , -2.5 < T-score ≤ -1) and Osteoporosis (OP, T-score ≤ -2.5). However, BMD alone does not fully capture bone quality, which encompasses mineral composition, collagen integrity, and microstructural organization, all of which are essential for determining bone strength and fracture susceptibility

[2]. Consequently, there is a pressing need for advanced diagnostic tools that provide comprehensive, molecular-level insights into bone health to improve early detection and management of these conditions.

Raman spectroscopy provides molecular information about bone minerals and organic components by quantifying inelastic scattering peaks as a function of wavenumber shift [3,4]. Metrics include several ratios of phosphate to collagen-associated bands, commonly termed mineral-to-matrix ratios (e.g., $PO_4^{3-}$/Amide I, $PO_4^{3-}$/Amide III, $PO_4^{3-}$/$CH_2$) as well as carbonate substitution, crystallinity, and collagen cross-linking, all of which relate to bone mechanical properties and disease states [5–7]. These parameters have been shown to correlate with bone strength, offering insights into how bone composition changes with aging, disease, and therapeutic interventions [8]. Spatially offset Raman spectroscopy (SORS) is a variant that enables non-invasive interrogation of subsurface layers in turbid media, including biological tissues [9]. Unlike conventional Raman spectroscopy, SORS spatially separates the laser illumination and Raman scatter collection points, allowing for depth-sensitive measurements that reduce contributions from surface layers [10]. While SORS has been widely explored for transcutaneous bone analysis, enabling the assessment of bone composition through overlying soft tissues [8,11–13], it can also provide depth-sensitive information about chemically different bone regions from a single excised bone, making it particularly valuable for controlled research settings where spectroscopic data can be directly correlated with histological and biomechanical analyses [14,15]. Recent studies have underscored the potential of SORS in assessing bone matrix quality and detecting differences associated with disease states. For instance, Rekha Gautam et al. (2023) demonstrated that SORS could detect changes in bone matrix quality in cadaveric bone samples subjected to autoclaving, a process that mimics disease-related alterations, showing sensitivity to changes in the mineral-to-matrix ratio and other Raman parameters [16]. In the context of osteoporosis, several studies have demonstrated how Raman spectroscopy can be used to distinguish between WHO diagnostic categories. For example, osteoporotic bone has been shown to exhibit lower phosphate-to-collagen (mineral-to-matrix) ratios and altered carbonate-to-phosphate ratios compared to healthy bone, reflecting changes in mineralization that impact bone quality [17–21].

For future clinical translation, particularly for transcutaneous Raman spectroscopy of the human finger, it will be essential to establish a reliable anatomical reference site for consistent comparison across individuals. Identifying and validating such a reference location requires comprehensive analysis on excised human cadaver fingers to ensure that Raman spectral markers can reliably distinguish between WHO bone-health categories. Establishing this reference framework in controlled cadaveric studies is a critical step toward enabling accurate and reproducible in vivo measurements in intact human subjects.

In this study, we apply SORS to excised human cadaver finger bones to classify them into normal, osteopenic, and osteoporotic categories based on their Raman spectral characteristics. By utilizing SORS's depth-resolved capabilities, we aim to capture detailed biochemical information about bone composition, including mineral-to-matrix ratios, carbonate substitution, and collagen integrity, to develop a non-destructive method for assessing bone health. Our approach not only enhances our understanding of the molecular underpinnings of bone diseases but also lays groundwork for potential clinical applications in the diagnosis and management of osteoporosis and osteopenia. Furthermore, by focusing on excised bone, in future studies we can directly correlate spectroscopic data with radiographic and biomechanical analyses, providing a comprehensive assessment of bone quality.

## 2. Materials and methods

### 2.1. Cadaver samples and bone preparation

Cadaveric finger bones from 25 female donors were obtained through the Anatomy Gifts Registry (Hanover, MD, USA). Inclusion criteria for donors included HIV-negative serology, less than 3-year postmortem recovery, and available medical history. Exclusions included known arm fractures, nonambulatory for more than 1-year, musculoskeletal impairments (paralysis or paresis), neoplasm, and prosthetic hardware. Donor characteristics including age, body mass index (BMI), ethnicity, 1/3 radius BMD, and T-score are summarized in Table 1. The cohort had a mean ± SD age of 70 ± 15 years (range 40–99 years) and was predominantly White (24 White, 1 Black). DXA acquired by Advanced Radiology (Hanover, MD, USA) using a Horizon Ci DXA System (Hologic) of the distal 1/3 radius classified n=8 donors as normal, n=6 as osteopenia, and n=11 as osteoporosis.

All cadavers were stored at −80 °C on arrival. Prior to collecting Raman measurements, exposed bone specimens were thawed and rehydrated for 2 h in 1x phosphate-buffered saline (PBS) at room temperature. Specimens were then wrapped in 1x PBS–soaked gauze to maintain hydration until Raman acquisition occurred. Upon completion of Raman acquisition, specimens were returned to the −80 °C freezer for future analyses.

Table 1. Demographics of cadaver donors (mean ± SD).

| WHO Class. | Ethnicity | Age | BMI | Distal 1/3 Radius BMD | 1/3 Radius T-score |
|---|---|---|---|---|---|
| | (Black/White) | (years) | (kg/m$^2$) | (g/cm$^2$) | |
| Normal | 1 / 7 | 56.5 ± 10.1 | 31.4 ± 6.4 | 0.698 ± 0.028 | 0.05 ± 0.45 |
| Osteopenia | 0 / 6 | 69.0 ± 7.6 | 24.7 ± 5.3 | 0.556 ± 0.023 | −1.63 ± 0.38 |
| Osteoporosis | 0 / 11 | 82.5 ± 9.9 | 21.2 ± 4.2 | 0.443 ± 0.054 | −4.19 ± 0.90 |

*2.2. SORS setup and Raman acquisition protocol*

In our setup, emitted light is imaged at 0, 3, and 6 mm from the laser excitation spot, linearly along the axis of the bone. This configuration captures SORS signals corresponding to progressively greater sampling depths, as illustrated conceptually in Fig. 1A, where the 3- and 6-mm photon paths indicate deeper light penetration at larger spatial offsets. Raman spectra were collected ex vivo from thawed, dissected finger bones after removal of surrounding soft tissue. To minimize anatomical variability and facilitate cross-cadaver comparison, every measurement was taken on the proximal phalanx of second digit (D2P1). As illustrated in Fig. 1B, the total length of the phalanx was measured, its midpoint marked (labeled M0), and the 0-mm offset fiber bundle was centered on this mark. At each site, we acquired five consecutive frames, 60 s each, and averaged them to yield a single high-signal-to-noise Raman spectrum per offset distance. The optical arrangement collected spatially offset Raman spectra from excised finger bones as illustrated in Fig. 1C. A continuous-wave 830-nm diode laser (Model PI-ECL-830-500-FS, Process Instruments Inc., Salt Lake City, UT) delivering 100 mW is directed onto the cortical surface through a 50-mm focal-length objective (numerical aperture ≈ 0.29). The beam strikes the bone at an angle of about 60° to the surface normal, producing a mildly elliptical surface spot approximately 1 mm along the ellipse's minor axis. Scattered light is captured by another objective with 30-mm focal length (numerical aperture ≈ 0.39) normal to the cortical surface, and after passing the dichroic mirror, is captured by three fused-silica fiber bundles. The bundles are fixed in a custom-designed fiber holder so that their centers are conjugated to the 0-, 3-, and 6-mm locations at the sample surface. Bundle A (capturing the brightest signal at 0 mm) consisted of four fibers, Bundle B of twelve fibers, and Bundle C of twenty-six fibers; in this fashion the net signals from the three fiber bundles were of comparable strength. All fibers had a 100 μm core, 120 μm cladding, and 140 μm buffer. To guarantee that the mechanical offsets correspond to true optical offsets on the curved bone surface, the probe tip is first back-illuminated, the glowing bundle faces are imaged through the objective onto a small USB CMOS camera, and the fibers on the probe mount are adjusted until the laser guide beam lies exactly at the first bundle.

Raman-scattered light from the three collection fiber bundles was arranged into a single column at the entrance to a Kaiser Optical Systems HoloSpec VPT imaging spectrograph (f/1.8). The input adapter aligns the fiber line with the slit so that the three bundles are spatially separated along the detector vertical axis. Inside the spectrograph, a volume-phase holographic

(VPH) grating disperses the light onto a thermoelectrically cooled Andor iDus DU420A-BEX2-DD back-illuminated deep-depletion CCD (1024×256 pixels, 26 μm pitch, operated at –55 °C).

Specimens are held on a motorized XYZ translation stage with 10-μm Z-resolution for rapid refocus. The stage origin is defined at the dorsal side of the distal phalanx, and subsequent measurement points are referenced to this coordinate system so that every finger is sampled at comparable anatomical locations. A camera mounted coaxially with the collection objective supplies real-time images of the laser spot and fiber faces, allowing fine focus adjustments and confirming that the offsets remain accurate despite local surface curvature.

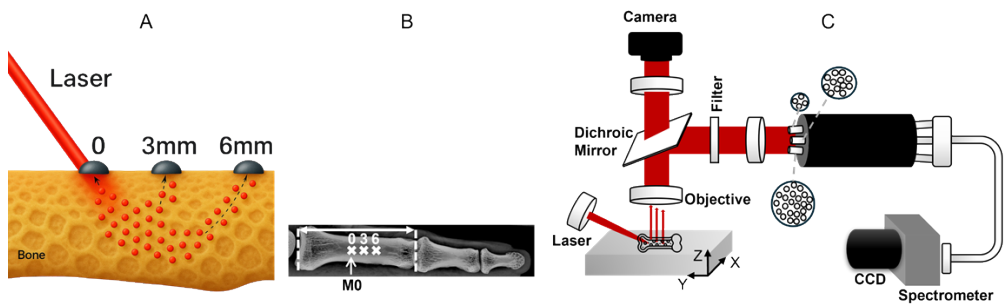

Fig. 1. SORS setup used in our experiment. A) Illustration of the SORS concept for excised bone measurements. Here, the detection fibers are positioned at offsets of 0, 3, and 6 mm from the laser illumination spot. As photons travel along "banana-shaped" paths from source to detector within the tissue, greater offset distances enable deeper penetration. B) Measurement geometry. The midpoint (M0) of each excised proximal phalanx of the second digit was located with a millimeter ruler. The excitation laser was centered on M0 and collected by the first fiber bundle (designated the 0-mm spatial offset). Additional offsets of 3 mm and 6 mm were obtained by translating the collection bundle distally along the bone. C) Optical setup for SORS used in this study. Three fiber bundles, containing four, twelve, and twenty-six fibers respectively, are physically spaced apart on the probe end so that, once aligned using back-illumination, they collect Raman signals at 0-, 3-, and 6-mm offsets (4, 12, and 26 fibers, respectively) from the laser spot on the sample. The sample is placed on a motorized translation stage (X, Y, Z axes) for precise positioning, and a camera helps verify focus and alignment.

### 2.3. Data analysis

All spectra were processed in MATLAB® R2025b with an in-house GUI. Raw 1024 × 256-pixel CCD frames (5 accumulations) were first dark-subtracted. Spatial and spectral aberrations were removed by fitting second-order polynomials to the centroids of 15 neon and 13 acetaminophen calibration peaks, yielding rectified images that were cropped (499 pixels). Pixel–wavelength calibration was obtained using a neon gas lamp; the Raman-shift axis was calibrated to acetaminophen to solve for the effective laser wavelength. Detector quantum efficiency and fiber throughput were normalized in two steps: (i) a high-frequency fixed-pattern derived from a broadband white-lamp image and (ii) a broadband spectral response derived from NIST-traceable green-glass fluorescence (ISRM-2241). Rows belonging to each fiber bundle were summed to give one spectrum for the 0-, 3-, and 6-mm spatial offsets).

Cosmic rays were rejected if their intensities exceeded 10 median-absolute-deviations above the mean of the four lowest values among five consecutive frames. Fluorescence was removed with an iterative algorithm [22] (order 7, 10 iterations, convergence < 0.05%); negative intensities lower than the noise floor ($0 \pm 0.5\sigma_{noise}$) were clipped. For outcome modelling, a leave-one-out cross-validation (LOOCV) partial least-squares regression (PLSR) was implemented with the plsregress function in MATLAB. Processed spectra served as predictors, while distal-radius T-scores were responses. Model rank (1–9) was chosen in each iteration, and performance was quantified by the Pearson correlation coefficient ($r$) and the root-mean-square error of cross-validation ($RMSE_{CV}$). For each model rank, LOOCV was performed, and the optimal rank was selected as the one yielding the minimum $RMSE_{CV}$.

## 3. Results

### 3.1. Offset-Dependent Spectral Discrimination

Mean Raman spectra per WHO category collected at the D2P1-M0 position for all cadavers at spatial offsets of 0 mm and 3 mm from the laser irradiation spot are depicted at Fig. 2A and Fig. 2B correspondingly. All spectra are normalized to the phosphate peak (≈ 960 cm$^{-1}$). In Fig. 2A, the mean spectra of normal and osteopenia are nearly indistinguishable. When 3-mm offset spectra are used, the mean spectra are more apparent between classes. The carbonate band increases significantly in the osteoporotic group compared to the normal healthy bone, while the Amide III and Amide I bands become notably stronger, indicating a higher collagen content relative to the normalized phosphate signal. The scatter plots in Fig. 3 and Fig. 4 quantitatively illustrate these trends across individual cadavers for the 0- and 3-mm offsets, respectively, with all corresponding numeric ratios summarized in Table 2. At 0 mm, only one mineral-to-matrix ratios show significant differences between normal and osteopenic bone. Comparisons between normal and osteoporotic bone, 4 ratios were significant but one was not. The osteopenia–osteoporosis (OPE–OP) comparison reveals significance for only 2 ratios. Overall, the limited statistical contrast underscores the shallow sampling depth achieved without spatial offset.

Introducing a 3-mm lateral offset markedly enhances diagnostic separation between groups as shown in Fig. 4. All five ratios now distinguish normal from osteoporotic bone with high significance (p ≤ 0.0001). Four ratios also differentiate normal from osteopenia (e.g., $PO_4^{3-}/CO_3^{2-}$, p = 0.007), while $PO_4^{3-}$/Amide I remain nonsignificant (p = 0.1555). The OPE–OP contrast is also strengthened at 3 mm, with four ratios achieving statistical significance (p ≤ 0.05). The gain likely reflects deeper photon trajectories that sample the subcortical region while retaining sufficient phosphate signal for reliable normalization. These findings confirm that increasing the spatial offset improves sensitivity to depth-dependent biochemical variations associated with bone demineralization.

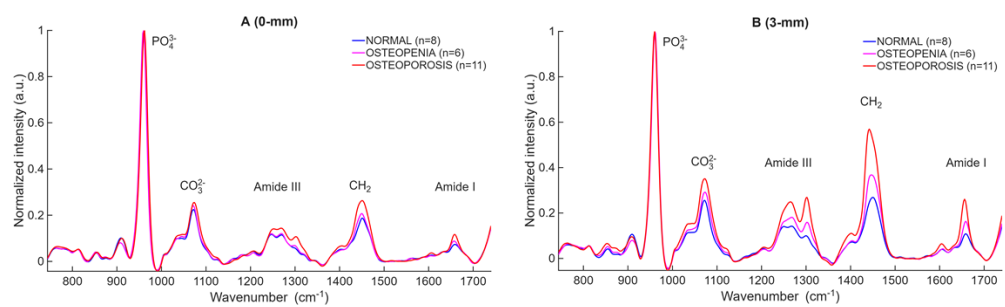

Fig. 2. Mean Raman spectra grouped by WHO category, healthy (n = 8), osteopenia (n = 6), osteoporosis (n = 11), spectra from midshaft (M0) of D2P1. (A) shows spectra acquired with the 0-mm collection bundle; (B) shows spectra acquired with a 3-mm lateral offset. Each spectrum is normalized to the $PO_4^{3-}$ peak at 960 cm$^{-1}$. Key matrix-sensitive bands $CO_3^{2-}$, Amide III, $CH_2$, and Amide I are labelled. The 3-mm offset reveals pronounced separation between normal (blue), osteopenia (magenta), and osteoporosis (red), whereas the 0-mm normal and osteopenia spectra overlap almost completely.

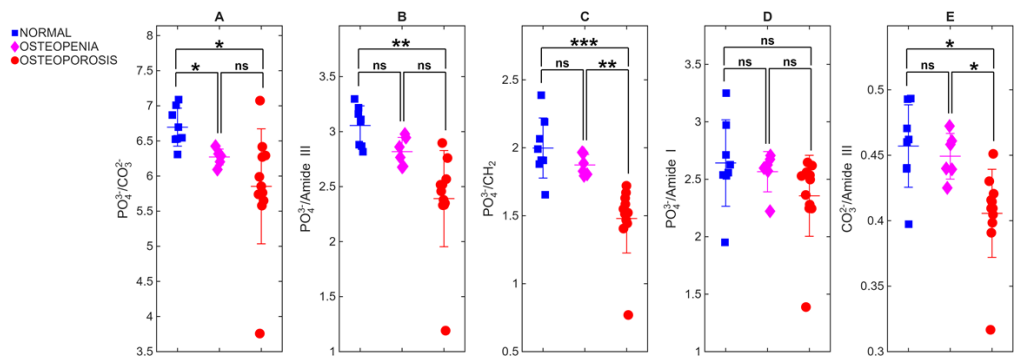

Fig. 3. Bone-specific Raman ratios at 0-mm spatial offset. Scatter plots display five diagnostic ratios. (A) $PO_4^{3-}/CO_3^{2-}$, (B) $PO_4^{3-}$/Amide III, (C) $PO_4^{3-}/CH_2$, (D) $PO_4^{3-}$/Amide I, and (E) $CO_3^{2-}$/Amide III for each cadaver (n=25). At 0 mm, only $PO_4^{3-}/CO_3^{2-}$ reach significance between normal and osteopenic bone, whereas $PO_4^{3-}/CH_2$ and $CO_3^{2-}$/Amide III discriminate osteopenia from osteoporosis, underscoring the limited diagnostic contrast without a spatial offset. *P*-values were annotated as follows: $p \leq 0.0001$ (****), $p \leq 0.001$ (***), $p \leq 0.01$ (**), $p \leq 0.05$ (*), and ns = not significant.

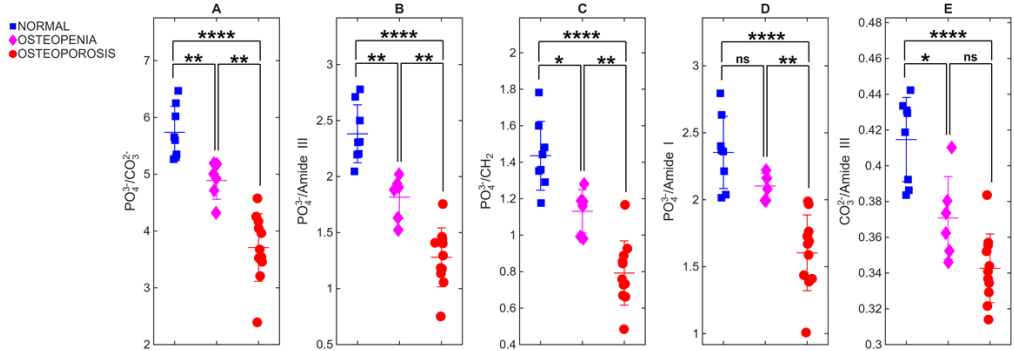

Fig. 4. Bone-specific Raman ratios at 3-mm spatial offset. The same five ratios as in Fig. 3 are plotted for spectra acquired at a 3-mm lateral offset. All ratios now separate normal from osteoporosis with high significance ($p \leq 0.0001$), and four ratios also differentiate normal from osteopenia and osteopenia from osteoporosis. The improved discriminability demonstrates that a 3-mm offset enhances sensitivity to depth-dependent biochemical changes while preserving sufficient phosphate signal for reliable normalization.

Table 2. Pairwise *p*-values for Raman band ratios comparing WHO-defined bone health categories at 0- and 3-mm spatial offsets (n = 25). P values ≤ 0.05 are indicated by bold font, and * show significance level.

| Ratio | Offset (mm) | p(N-OP) | p(N-OPE) | p(OPE-OP) |
|---|---|---|---|---|
| $PO_4^{3-}/CO_3^{2-}$ | 0 | **0.0386*** | **0.0113*** | 0.7186 |
|  | 3 | **≤0.0001****** | **0.0072**** | **0.0014**** |
| $PO_4^{3-}$/Amide III | 0 | **0.0026**** | 0.0524 | 0.1073 |
|  | 3 | **≤0.0001****** | **0.0023**** | **0.0016**** |
| $PO_4^{3-}/CH_2$ | 0 | **0.0007***** | 0.6444 | **0.0070**** |
|  | 3 | **≤0.0001****** | **0.0145*** | **0.0023**** |
| $PO_4^{3-}$/Amide I | 0 | 0.3234 | 1.0000 | 0.5860 |
|  | 3 | **≤0.0001****** | 0.1555 | **0.0026**** |
| $CO_3^{2-}$/Amide III | 0 | **0.0107*** | 1.0000 | **0.0301*** |
|  | 3 | **≤0.0001****** | **0.0139*** | 0.0504 |

To determine whether a greater offset further improves diagnostic power, we compared spectra collected at 3 mm and 6 mm from a subset of 13 cadaveric fingers (n=5 normal, n=5 osteopenia, and n=3 osteoporotic) from which we had both 3- and 6-mm offsets. Fig. 5 contrasts the performance of these two spatial offsets. At 3 mm (Fig. 5A), the mean spectra for normal, osteopenia, and osteoporosis separate visibly at several matrix-sensitive bands ($CO_3^{2-}$, Amide III, $CH_2$, Amide I), using the phosphate peak for normalization. The three categories remain well separated across all matrix-sensitive bands, and the corresponding ratio plots retain strong statistical significance for every metric (Table 3). All five ratios discriminate normal from osteoporotic bone ($p \leq 0.05$), and four ratios also differentiate normal from osteopenic bone and three ratios differentiate osteopenia from osteoporosis. When the offset is increased to 6 mm (Fig. 5B), the overall trends between groups remain the same, but the separations become less pronounced and data variability increases, reducing classification strength. All five ratios remain significant for normal vs. osteoporosis. For normal vs. osteopenia, three ratios remain significant ($p \leq 0.05$), whereas $PO_4^{3-}/CH_2$ and $PO_4^{3-}$/Amide I fail to reach significance. The osteopenia–osteoporosis comparison also weakens, with only $PO_4^{3-}/CH_2$ and $PO_4^{3-}$/Amide I remain significant. Compared with the 3-mm offset, these p-values indicate that the 6-mm does not enhance discrimination between different groups compared with 3-mm offset.

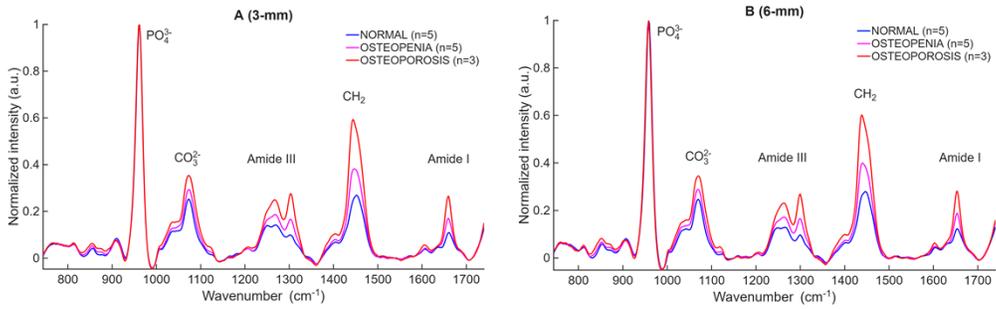

Fig. 5. Representative mean Raman spectra for the cohort of 13 cadaveric fingers, healthy (n = 5), osteopenia (n = 5), osteoporosis (n = 3) spectra from the midshaft (M0) of D2P1, illustrating the effect of increasing the spatial offset. (A) shows the spectra collected with a 3-mm lateral offset; (B) shows the spectra collected with a 6-mm offset. Each spectrum is normalized to the $PO_4^{3-}$ peak at 960 cm$^{-1}$.

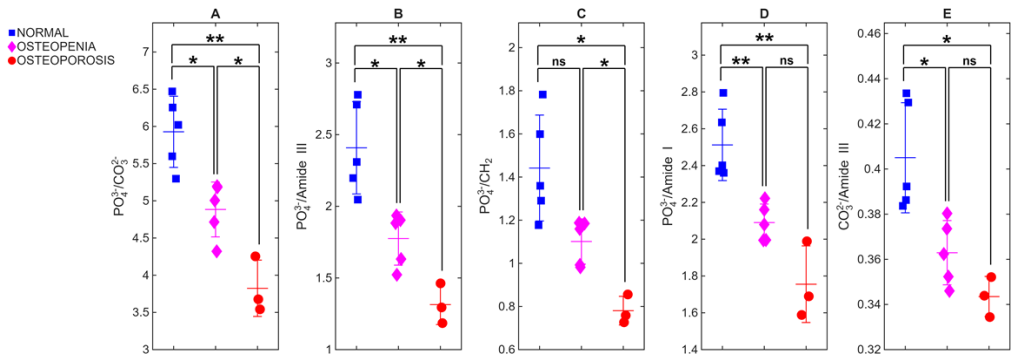

Fig. 6. Raman-derived compositional ratios for the same cohort as Fig. 5 at 3-mm offset. Scatter plots (A–E) display $PO_4^{3-}/CO_3^{2-}$, $PO_4^{3-}$/Amide III, $PO_4^{3-}/CH_2$, $PO_4^{3-}$/Amide I, and $CO_3^{2-}$/Amide III, respectively, for each cadaver (n = 13). All five ratios significantly differentiate normal from osteoporosis, with several also distinguishing normal from osteopenia and osteopenia from osteoporosis.

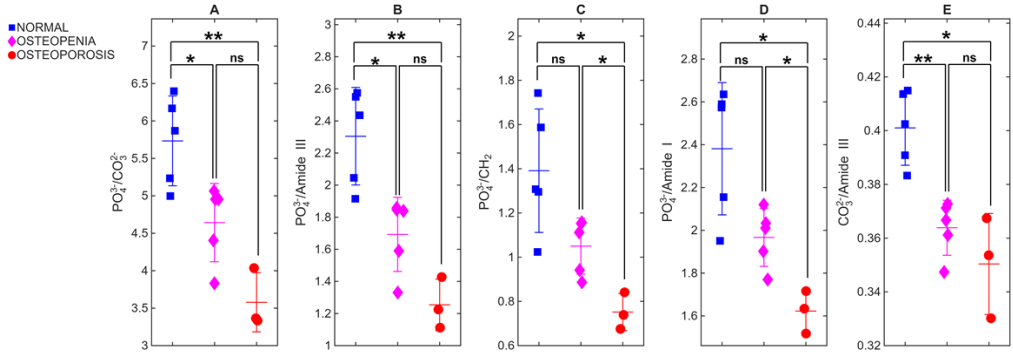

Fig. 7. Raman-derived compositional ratios for the same cohort as Fig. 5 at a 6-mm offset. The same ratios as in Fig. 6 are plotted for spectra acquired at a 6-mm lateral offset. All five ratios remain significant for normal vs osteoporosis. For normal vs osteopenia, three ratios retain significance (p ≤ 0.05) and only two ratios, $PO_4^{3-}/CH_2$ and $PO_4^{3-}$/Amide I, distinguish osteopenia from osteoporosis.

Table 3. Pairwise *p*-values for Raman band ratios comparing WHO- defined bone health categories at 3- and 6-mm spatial offsets (n = 13). P values ≤ 0.05 are indicated by bold font, and * show significance level.

| Ratio | Offset (mm) | p(N-OP) | p(N-OPE) | p(OPE-OP) |
| --- | --- | --- | --- | --- |
| $PO_4^{3-}/CO_3^{2-}$ | 3 | **0.0020**** | **0.0142*** | **0.0239*** |
|  | 6 | **0.0047**** | **0.0462*** | 0.0707 |
| $PO_4^{3-}$/Amide III | 3 | **0.0047**\*\* | **0.0154**\* | **0.0312**\* |
| | 6 | **0.0048**\*\* | **0.0216**\* | 0.0863 |
| $PO_4^{3-}$/$CH_2$ | 3 | **0.0132**\* | 0.0642 | **0.0102**\* |
| | 6 | **0.0285**\* | 0.1136 | **0.0354**\* |
| $PO_4^{3-}$/Amide I | 3 | **0.0060**\*\* | **0.0077**\*\* | 0.0594 |
| | 6 | **0.0209**\* | 0.0753 | **0.0272**\* |
| $CO_3^{2-}$/Amide III | 3 | **0.0191**\* | **0.0309**\* | 0.2443 |
| | 6 | **0.0135**\* | **0.0040**\*\* | 0.6814 |

Considering all 25 samples, among the five metrics, $PO_4^{3-}$/$CO_3^{2-}$ and $PO_4^{3-}$/Amide III emerge as the most robust indicators, for Normal–Osteoporosis, Osteopenia–Osteoporosis as well as Osteopenia–Osteoporosis comparison, at the 3-mm offset.

### 3.2. T-Score Prediction by PLS Regression

To explore whether Raman spectra recorded on the phalanges can serve as a surrogate for wrist bone mineral health, we built PLSR-LOOCV using both 0- and 3-mm spatial-offset measurements. The predictor matrix consisted of Raman spectra collected at the mid-shaft position (M0) of the D2P1 using both 0- and 3-mm spatial offsets. For the 3-mm offset data, the laser illuminated the M0 site, while the detector was placed 3 mm distally. The response vector comprised the DXA T-scores of the distal radius for 25 samples. Fig. 8 summarizes the outcome for both the 0- and 3-mm configuration. In Fig. 8A, each symbol represents one cadaver; the measured T-score (abscissa) is plotted against the T-score predicted solely from the Raman spectra (ordinate). With the 0-mm dataset, the model achieved a correlation of r = 0.678 and a $RMSE_{CV}$ = 1.48 T-score units. Although the correlation trend is correct, a substantial number of predictions cross the WHO diagnostic thresholds, resulting in frequent band misclassifications. Replacing the predictors with the 3-mm-offset spectra markedly improved performance. As shown in Fig. 8B, the correlation increased to r = 0.85, while the error decreased to $RMSE_{CV}$ = 1.044 units. 92% of specimens now fall within their correct diagnostic categories, and the few misclassifications lie within the ±1.044 uncertainty implied by the $RMSE_{CV}$.

The corresponding grouped data (Fig. 8C and Fig. 8D) provide a complementary view of these results. At 0 mm, predicted (teal circles) and measured (orange squares) T-scores show broad overlap across the normal and osteopenic groups. In contrast, at 3 mm, the median predicted T-score for each cohort (teal circles) closely tracks the measured median (orange squares), typically within one T-score unit. This close agreement underscores the improved discriminative accuracy and clinical relevance of the spatial-offset predictions.

The correlation between measured and predicted T-scores and the reduced prediction error demonstrates that depth-sensitive Raman spectroscopy can estimate the DXA T-score with clinically meaningful accuracy. Taken together, these results show that introducing a 3-mm spatial offset, sufficient to probe the subcortical bone region while maintaining a reliable phosphate reference, nearly halves the prediction error relative to 0-mm collection and achieves diagnostic accuracy levels suitable for pre-screening of bone health.

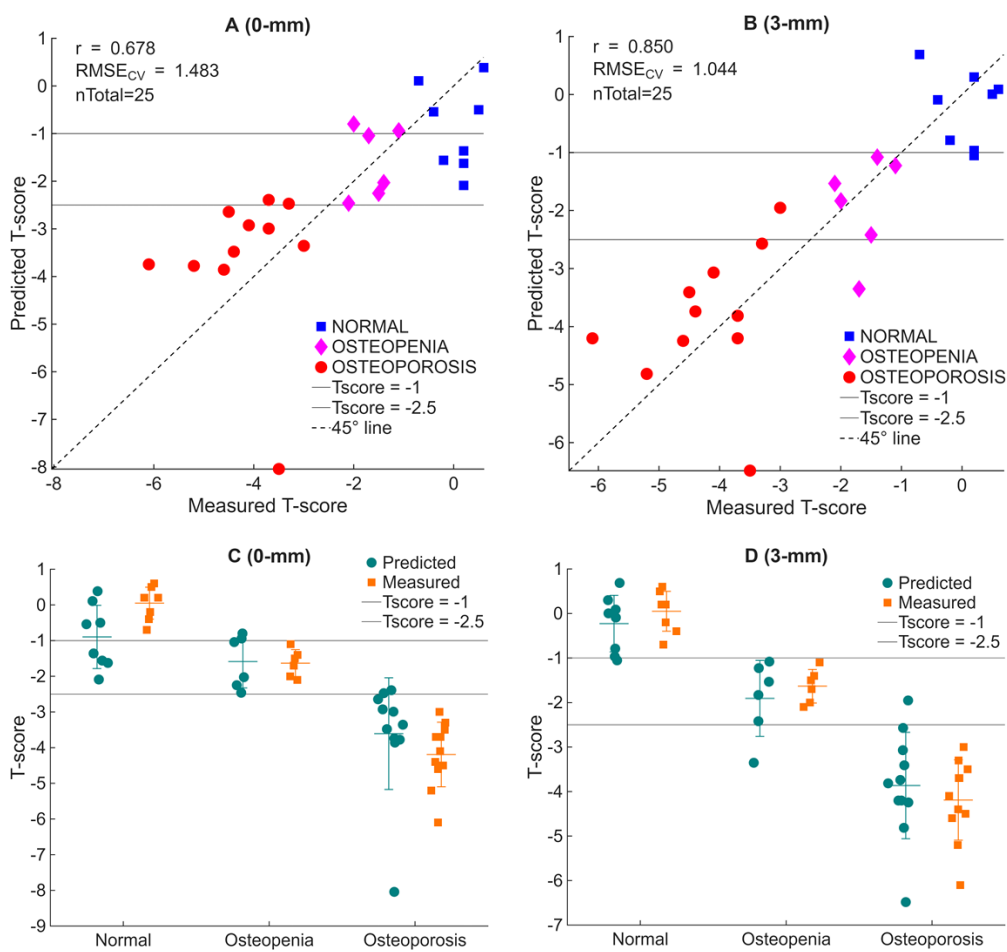

Fig. 8. Distal radius DXA T-score predicted from phalangeal Raman spectra using PLSR with leave-one-out cross-validation (n = 25). Grey horizontal lines mark the WHO thresholds at T = -1 and T = -2.5; the dashed line is the 45° identity. (A) Measured versus predicted T-scores form the 0-mm Raman spectra (r = 0.678, RMSECV = 1.483). (B) Measured versus predicted T-scores from 3-mm spatial-offset spectra (r = 0.850, RMSE$_{CV}$ = 1.044)**,** showing improved correlation and reduced error. (C–D) Same data grouped by diagnostic category. Predicted (teal circles) and measured (orange squares) T-scores are shown with medians and standard deviations. At 3 mm, median predicted values approximate measured medians within 1 T-score unit, illustrating enhanced discriminative accuracy at the spatial offset.

## 4. Discussion

Our findings confirm that SORS can interrogate subsurface regions of small bones, such as the phalanges, and extract composition-dependent markers that correlate with clinically established indicators of bone health, namely WHO diagnostic categories using the 1/3 radius DXA T-scores. The strongest separation between normal, osteopenic, and osteoporotic bone was achieved at a 3-mm lateral offset, where all five ratios significantly differentiated normal from osteoporotic and four of the five mineral-to-matrix ratios showed significant differences between normal and osteopenic bone and between osteopenic and osteoporotic bone, confirming that biochemical contrast increases with offset. This enhanced discriminability at 3 mm is likely due to the offset's ability to sample the subcortical bone region, which is more sensitive to mineral and matrix alterations and compositional changes associated with osteoporosis [2,14]. In contrast, the 0-mm configuration, which predominantly probes the superficial cortex, yielded only one significant ratio ($PO_4^{3-}/CO_3^{2-}$) for normal vs osteopenia and two ratios ($PO_4^{3-}/CH_2$, and $CO_3^{2-}/Amide\ III$) for osteopenia vs osteoporosis, underscoring the limited diagnostic sensitivity of surface-weighted Raman collection. Together, these results highlight the critical role of depth-resolved SORS measurements in detecting compositional changes indicative of bone demineralization and structural degradation during the onset of osteoporosis. These observations were obtained from ex vivo phalanges with the excised bone; within this geometry, offsets in the 3–6 mm range appear suitable. Importantly, optimal offsets for intact full-hand ex vivo or in-vivo measurements may differ because the photon origin is shifted toward the epithelial surface of the phalanx, and intervening soft tissue alters both SNR and sampling volume [13]. Future studies could employ computational models or experimental techniques to better characterize the sampling volumes associated with different offsets in phalangeal bone.

Compared with our group's closest prior work [19], in which wrist DXA T-scores were modeled from excised proximal phalanx spectra using LOOCV-PLSR on $n = 12$ cadaver forearms, the present 3-mm SORS model demonstrates superior performance. In the previous study, we reported a correlation of $r = 0.60$ with $RMSE_{CV} = 1.46$ for diaphyseal spectra. In contrast, in the current work, the 3-mm offset model achieved $r = 0.85$ with $RMSE_{CV} = 1.04$ across $n = 25$ samples and correctly classified 92% of specimens into their WHO diagnostic categories, with the few remaining misclassifications falling within the ±1 T-score uncertainty implied by the $RMSE_{CV}$. While the experimental designs differ, our previous work employed measurements without a spatial offset, whereas the present study incorporated a 3-mm offset, the higher correlation and lower error observed here highlight the depth-sensitivity advantage of SORS. These results underscore that depth-resolved phalangeal Raman spectra can serve as a reliable surrogate for wrist bone mineral health, offering clinically meaningful predictive accuracy without using ionizing radiation.

The ex vivo design, while allowing for controlled measurements, does not account for soft-tissue attenuation and motion artifacts encountered in vivo. However, prior research has demonstrated the feasibility of SORS in vivo, suggesting that with appropriate adjustments, such as enhanced signal processing or probe optimization, this technique could be translated to clinical settings [11,12]. The high classification accuracy and strong correlation with wrist DXA T-scores highlight the potential of SORS for fast, non-ionizing pre-screening for osteoporosis. Future in vivo studies are essential to validate these findings and address challenges such as soft-tissue interference and patient motion, paving the way for practical implementation of SORS as a prescreening tool.

## 5. Conclusion

Spatially offset Raman spectroscopy (SORS) effectively captures depth-dependent biochemical variations in excised human finger bones and differentiates normal, osteopenic, and osteoporotic specimens based on ratios. Compared with 0-mm offset collection, a 3-mm offset markedly improved bone-health classification fidelity and enabled a PLSR model to predict distal-radius DXA T-scores with $RMSE_{CV} \approx \pm 1$ and a correlation of $r = 0.85$. Increasing the offset to 6 mm did not improve diagnostic contrast. These results demonstrate that subsurface Raman measurements can serve as a nonionizing surrogate for DXA-derived bone metrics. Future work should extend this approach to transcutaneous and in-vivo measurements, advancing Raman spectroscopy toward deployment as a rapid, non-ionizing screening tool for bone-health assessment in both primary-care and resource-limited settings.

**Funding.** The study was supported by grant numbers R01AR070613 (AB and HA), and P30AR069655 (HA) from NIAMS/NIH. The content is solely the responsibility of the authors and does not necessarily represent the official view of the NIH.

**Acknowledgments.** The authors would like to thank Lindsay Schnur for her technical assistance with μCT.

**Disclosures.** The authors declare that they have no known competing financial interests or personal relationships that could have appeared to influence the work reported in this paper.

**Data Availability.** Data underlying the results presented in this paper are available from the corresponding authors upon reasonable request.

## References

1. J. Kanis, C. Cooper, R. Rizzoli, and J.-Y. Reginster, "European guidance for the diagnosis and management of osteoporosis in postmenopausal women," Osteoporos. Int. **30**, 3–44 (2019).
2. E. Seeman and P. D. Delmas, "Bone Quality — The Material and Structural Basis of Bone Strength and Fragility," New Engl. J. Med. **354**, 2250–2261 (2006).
3. G. S. Mandair and M. D. Morris, "Contributions of Raman spectroscopy to the understanding of bone strength," BoneKEy Reports **4** (2015).
4. M. D. Morris and G. S. Mandair, "Raman Assessment of Bone Quality," Clin. Orthop. & Relat. Res. **469**, 2160–2169 (2011).
5. M. Khalid, T. Bora, A. A. Ghaithi, *et al.*, "Raman Spectroscopy detects changes in Bone Mineral Quality and Collagen Cross-linkage in Staphylococcus Infected Human Bone," Sci. Reports **8**, 9417 (2018).
6. M. Unal, R. Ahmed, A. Mahadevan-Jansen, and J. S. Nyman, "Compositional assessment of bone by Raman spectroscopy," The Anal. **146**, 7464–7490 (2021).
7. S. Fertaki, P. Giannoutsou, and M. G. Orkoula, "Combining Raman Microspectroscopy and X-ray Microcomputed Tomography for the Study of Bone Quality in Apolipoprotein-Deficient Animal Models," Molecules **28**, 7196 (2023).
8. K. Buckley, J. G. Kerns, J. Vinton, *et al.*, "Towards the in vivo prediction of fragility fractures with Raman spectroscopy," J. Raman Spectrosc. **46**, 610–618 (2015).
9. P. Matousek, I. P. Clark, E. R. C. Draper, *et al.*, "Subsurface Probing in Diffusely Scattering Media Using Spatially Offset Raman Spectroscopy," Appl. Spectrosc. **59**, 393–400 (2005). Publisher: Society for Applied Spectroscopy.
10. S. Mosca, C. Conti, N. Stone, and P. Matousek, "Spatially offset Raman spectroscopy," Nat. Rev. Methods Primers **1**, 21 (2021).
11. P. Matousek, E. R. C. Draper, A. E. Goodship, *et al.*, "Noninvasive Raman Spectroscopy of Human Tissue in vivo," Appl. Spectrosc. **60**, 758–763 (2006).
12. C. Shu, K. Chen, M. Lynch, *et al.*, "Spatially offset Raman spectroscopy for in vivo bone strength prediction," Biomed. Opt. Express **9**, 4781 (2018).
13. K. Chen, C. Massie, H. A. Awad, and A. J. Berger, "Determination of best Raman spectroscopy spatial offsets for transcutaneous bone quality assessments in human hands," Biomed. Opt. Express **12**, 7517 (2021).
14. K. Sowoidnich, J. H. Churchwell, K. Buckley, *et al.*, "Spatially offset Raman spectroscopy for photon migration studies in bones with different mineralization levels," The Anal. **142**, 3219–3226 (2017).
15. Z. Liao, F. Sinjab, A. Nommeots-Nomm, *et al.*, "Feasibility of Spatially Offset Raman Spectroscopy for in Vitro and in Vivo Monitoring Mineralization of Bone Tissue Engineering Scaffolds," Anal. Chem. **89**, 847–853 (2017).
16. R. Gautam, R. Ahmed, E. Haugen, *et al.*, "Assessment of spatially offset Raman spectroscopy to detect differences in bone matrix quality," Spectrochimica Acta Part A: Mol. Biomol. Spectrosc. **303**, 123240 (2023).
17. B. R. McCreadie, M. D. Morris, T.-c. Chen, *et al.*, "Bone tissue compositional differences in women with and without osteoporotic fracture," Bone **39**, 1190–1195 (2006).
18. E. P. Paschalis, P. Fratzl, S. Gamsjaeger, *et al.*, "Aging Versus Postmenopausal Osteoporosis: Bone Composition and Maturation Kinetics at Actively-Forming Trabecular Surfaces of Female Subjects Aged 1 to 84 Years," J. Bone Miner. Res. **31**, 347–357 (2016).
19. C. Massie, E. Knapp, H. A. Awad, and A. J. Berger, "Detection of osteoporotic-related bone changes and prediction of distal radius strength using Raman spectra from excised human cadaver finger bones," J. Biomech. **161**, 111852 (2023).
20. K. Chen, C. Yao, M. Sun, *et al.*, "Raman spectroscopic analysis for osteoporosis identification in humans with hip fractures," Spectrochimica Acta Part A: Mol. Biomol. Spectrosc. **314**, 124193 (2024).
21. W. Yang, S. Xia, X. Jia, *et al.*, "Utilizing surface-enhanced Raman spectroscopy for the adjunctive diagnosis of osteoporosis," Eur. J. Med. Res. **29**, 476 (2024).
22. C. A. Lieber and A. Mahadevan-Jansen, "Automated Method for Subtraction of Fluorescence from Biological Raman Spectra," Appl. Spectrosc. **57**, 1363–1367 (2003).